\documentclass[rnote]{aa} 
\usepackage{graphicx}
\usepackage{txfonts}
\usepackage{longtable}
\usepackage{pdflscape}
\def \etal {\textit{et al. }}

\begin{document}
    \title{Position angles and coplanarity of multiple systems from transit timing}
    \author{Aviv Ofir \inst{1,2}}

    \institute{Institut f\"ur Astrophysik, Georg-August-Universit\"at, 
Friedrich-Hund-Platz 1, 37077 G\"ottingen, Germany. \and 
\email{avivofir@astro.physik.uni-goettingen.de}}

    \date{Received XXX; accepted YYY}


   \abstract
    {}
    {We compare the apparent difference in timing of transiting planets (or eclipsing binaries) that are observed from widely separated locations (parallactic delay).}
    {A simple geometrical argument allow us to show that the apparent timing difference depends also on the on-sky position angle of the planetary (or secondary) orbit, relative to the ecliptic plane.}
    {We calculate the magnitude of the effect for all currently known planets (should they exhibit transits), finding that almost 200 of them -- mostly radial-velocity detected planets -- have predicted timing effect larger than 1 second. We also compute the theoretical timing precision for the PLATO mission, that will observe a similar stellar population, and find that a 1 second effect would be frequently readily observable. We also find that on-sky coplanarity of multiple objects in the same system can be probed more easily than the on-sky position angle of each of the objects separately.}
    {We show a new observable from transit photometry becomes available when very high precision transit timing is available. We find that there is a good match between projected capabilities of the future space missions PLATO and CHEOPS and the new observable. We give some initial science question that such a new observable may be related to and help addressing.}

\titlerunning{Position Angles and Coplanarity Of Transiting Systems}

\maketitle
%

\section{Basic principles}

We consider a system of one or several eclipsing bodies. These components can be either eclipsing stars or transiting exoplanets, but in the text below we will use only transiting exoplanets as a specific example, since similar effects for eclipsing binaries are even larger. We then assume that a particular transit event on that system was observed from two remote locations simultaneously, and the following was developed with \emph{Kepler} spacecraft (Borucki \etal 2010) and an Earth-bound observer in mind - so the observers are separated by AU-scale distances.

While the existence of transits practically ensures that the inclination angle $i$ is close to $90^\circ$, the on-sky orientation of the systems remains unconstrained. We wish use the above observational configuration to put such constraints on each transiting member of the system, and if there are more than one - on their on-sky coplanarity.

\begin{figure}
\includegraphics[width=0.22\textwidth]{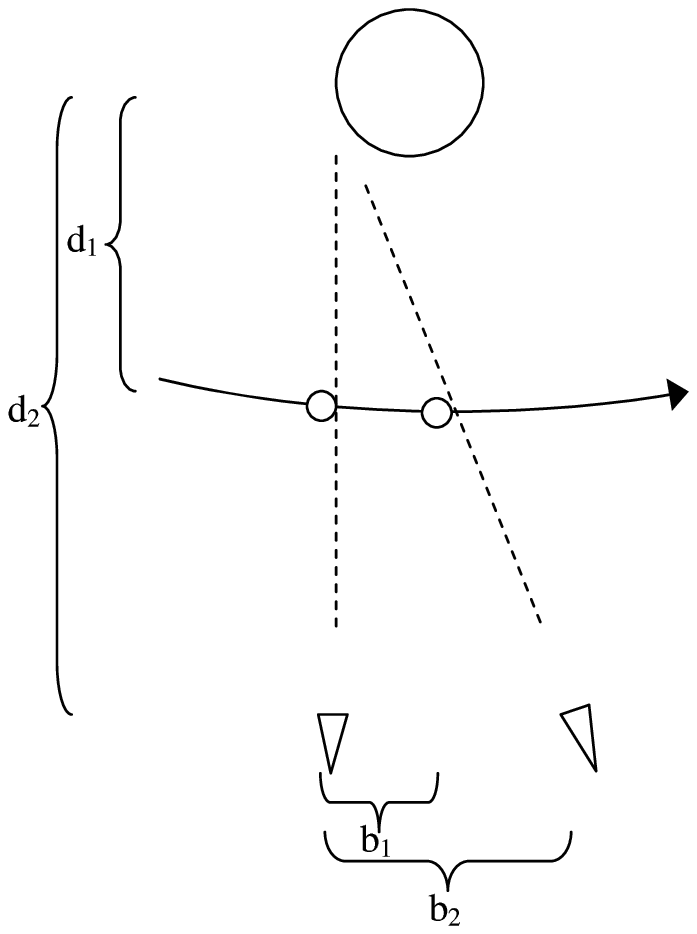}
\includegraphics[width=0.22\textwidth]{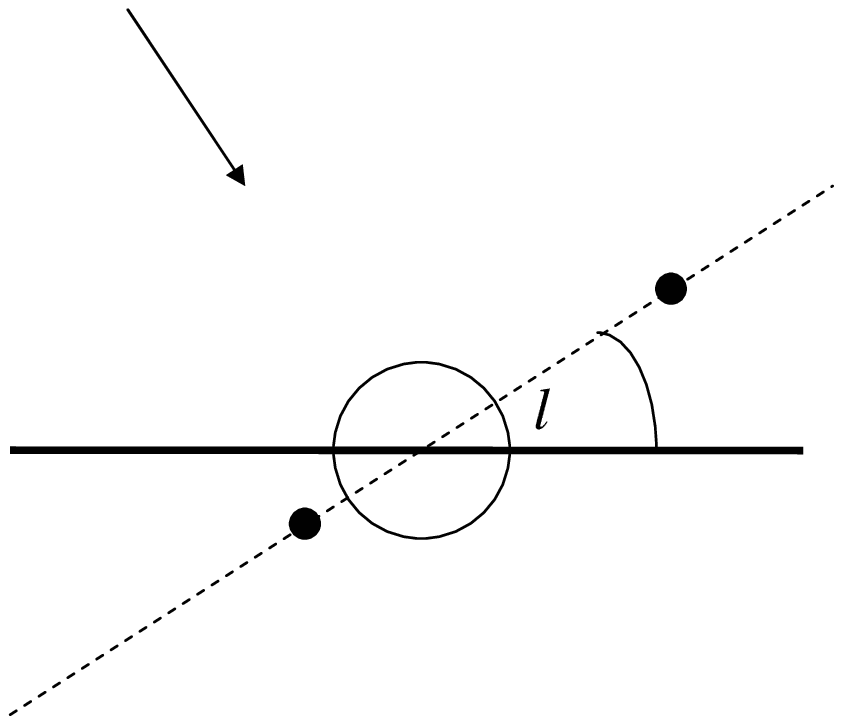}
\caption{\textbf{Left panel:} top view of the two observers (at the bottom) from above the plane containing the observers and the planet host star. As the planet (empty circle) moves in its orbit it will appear to transit (here the first contact is shown) at slightly different times for the two observers. \textbf{Right panel:} a view along the orbital plane of the planet from infinity. The two observers (filled circles) are on a line that is inclined at an angle \textit{l} relative to the orbital plane (the left panel view was taken from a position illustrated by the arrow).}
\label{Views}
\end{figure}

We begin by defining all the distances given in the left panel of figure \ref{Views}: the orbital distance during transit $d_1$, the distance to the system $d_2$, the baseline distance between the two observers $b_2$ and the corresponding distance projected on the planet's orbit $b_1$. Importantly, one expects that the planet will be seen by the different observers at the same transit phase (e.g., first contact is shown here) at a slight delay, also known as parallactic delay, of $D_{expect}=b_1/V$ where $V$ is the planet's orbital velocity $V=2 \pi d_1 / P$ (for circular orbits, where $P$ is the planet's orbital period), and $b_1=b_2 d_1 / d_2$, or:

\begin{equation}
D_{expect}=P \frac{b_2}{2 \pi d_2} = 0.06666 \left[\frac{P}{\mathrm{day}}\right]  \left[\frac{b_2}{\mathrm{AU}}\right]  \left[\frac{d_2}{\mathrm{pc}}\right]^{-1} [sec]
\label{No_l}
\end{equation}

Note the above can also be read simply as the planetary period times the parallax angle subtended by the two observers from the host star. Small but nonzero eccentricity $e$ would change $V$ by a fractional amount close to $e$ (i.e. $|\Delta V|/V \simeq e$), where the change can be either positive or negative, relative to the circular case, depending on the argument of periastron $\omega$. We note that higher order effects of finite ecentricity, such as orbital precession, are not currently included.

This delay is not related, and is actually perpendicular, to the light-time delay between the observers that is caused by different observer positions along the line of sight. If one adds the possibility that the planet's orbit may be tilted at an angle $l$ relative to the line separating the two observers (right panel of Figure \ref{Views}), then the effective separation between the two observers, as seen projected on the planet's orbit, is only $b_2 cos(l)$, so:
\begin{equation}
D_{observe}=P \frac{b_2}{2 \pi d_2} cos(l)
\end{equation}

One finds that by dividing $D_{observe}$ by $D_{expect}$ one can measure $cos(l)$, and this effect will be best seen for long period planets (if their period is known precisely enough) and host stars with higher parallactic angle. This measurement is different from the spectroscopic Rossiter-McLaughlin effect (e.g. Ohta et al. 2005) since it bares no relation to the stellar spin axis, which one may assess independently. 

Earth-bound observers travel significant distance yearly as they orbit the Sun (and a satellite in the L2 Lagrange point, such as PLATO (Rauer \& Catala 2011), is similar in that respect). Therefore, the global modeling of multiple transits taken by a \textit{single} observer can produce similar results to the two-observer model above as long as the prediction uncertainty of the linear ephemeris is small relative to the individual timing undertainty. In such case the parallactic delay effect will be manifested as residuals to the linear ephemeris that have a 1yr period and phase such that the maximal residuals occurs when the observer's orbit about the Sun, projected on a line perpendicular to the direction of the host star, is maximal. The two separate observations of the same transit are therefore not strictly required but are just more illustrative of the parallactic delay effect. Other timing effects, such as ones caused by orbital precession or dynamical interaction, do not have this particular morphology and can thus be disentangled from the parallactic delay.

In cases where accurate parallax is not known, $cos(l)$ may not be constrained in this way. However, even in these cases one may compare two eclipsing objects in the same system (when available) - e.g., two transiting planets - for their on-sky coplanarity even with no information on the system's distance:

\begin{equation}
\frac{cos(l_1)}{cos(l_2)}= \frac{D_1}{D_2} \frac{P_2}{P_1} \frac{b_{2,2}}{b_{2,1}} 
\end{equation}

Here the right side of the equation is made of only observable quantities, and the left side is constraining the on-sky angle between the two transiting planets. We note that at the timing precision required to observe this effect (see next section) the transit timing signal may also become sensitive to planet-planet interaction (Holman \& Murray 2005) - so these will need to be accounted for as well. We stress again that the parallactic delay effect has a very specific shape (a sine with a pre-determined period and pre-determined phase) and so it should be possible to disentangle this particular effect from other timing effects.

\section{Estimating signal significance}

\subsection{Known systems}
In order to assess the possibility of measuring this effect in already-known systems, we calculate $D_{expect}$ using eq. \ref{No_l} for all the planets on the NASA Exoplanet Archive\footnote{http://exoplanetarchive.ipac.caltech.edu/cgi-bin/ExoTables/nph-exotbls?dataset=planets  \quad\quad as of August 28, 2013} using $b_2=1$AU for all - had they were transiting their host star. For scale, we note that ground-based single-transit timing precision nowadays routinely reaches roughly 10 seconds, and sometimes even less than that (e.g. Tregloan-Reed \etal 2013). When sorting the $D_{expect}$ above, exoplanets that were already detected by direct imaging naturally turned out to be prime candidates (they are long-period and close to the Solar system) and these planets form all the top 3, and 14 of the top 19 largest signals (amplitudes of 19 seconds or more) -- but these planets are too long-period for practical follow-up, and their period is not constrained nearly as well as needed. This is unfortunate since such a configuration could have been use to independently validate this technique from the imaging data. 

More importantly, all the first 190 planets have $D_{expect}$ in excess of 1 second, which is not too far from the current state of the art, and most of these planets ($>90\%$) are planets detected by either radial velocity (RV) or transit (the latter of which means they were RV confirmed as well)- so they are relatively bright, helping to achieve good timing precision. Such planets, if found to be transiting, are prime candidates for constraining $cos(l)$. Of the planets that are already known to transit the most promising are HD 80606 b and Kepler-22 b, with $D_{expect}$ of 0.127s and 0.102s, respectively, so significantly more challenging.

\subsection{PLATO and CHEOPS}
In the following text we give the planned PLATO mission as a specific example since for this mission the predicted effect may be measured in large scale, even though isolated suitable cases may already be available in other data (above).

The GAIA mission is expected to measure the parallactic angle to all relevant targets to very high precision. Specifically, all PLATO host stars will be brighter than 11th magnitude \footnote{PLATO Definition Study Report (Red Book)}, while all stars brighter than 12th magnitude will be measured by GAIA to better than $14 \mu as$ \footnote{From GAIA web page}, so an 11th magnitude Sun like star will have a distance of $d_2 \approx 171.4$pc and have $d_2$ measured to $0.24\%$ or better. Furthermore, the relative errors on $P$, $b_2$ are typically several orders of magnitude smaller still, so while distances are notoriously difficult to measure in astronomy, GAIA will allow for the error on $D_{expect}$ to be negligible relative to the error in $D_{observe}$.

The times of mid-transit, and hence $D_{observe}$, can be measured with an accuracy of $ \sigma_{Tc} \simeq \left(t_e/2\Gamma\right)^{1/2}\sigma_{ph}\rho^{-2}$ (Ford \& Gaudi 2006) where is $t_e$ the duration of ingress/egress, $\Gamma$ is the rate at which observations are taken, $\sigma_{ph}$ is the photometric uncertainty and $\rho$ is the ratio of the planet radius to stellar radius. Since the minimum ingress or egress time is when the transits are central, and since the transit duty cycle for such transits at circular orbits is $q=\frac{R}{\pi a}$, the ingress/egress time $t_e$ is:

\begin{equation}
t_e=2P \, q \, \rho
\end{equation}

For the PLATO mission star samples P1 and P2 are required to have noise level $\leq 34$ ppm/hr, while for star sample P4 this requirement would be $\leq 80$ ppm/hr, and these stars would be observed for two to three years nearly continuously. Assuming white noise, one can scale a given noise level requirement with the ingress/egress time $t_e$ for a given orbital period, and compare the expected timing accuracy of single transits to that is expected by $D_{expect}$. This, however, would underestimate PLATO's detection capability since by observing multiple transits events (roughly $N_{tr}=\frac{2yr \textrm{ to } 3yr}{P}Q_{PL}$ events) one would gain a factor of $N_{tr}^{1/2}$ in detectable precision (where $Q_{PL}$ is the PLATO-required duty cycle of $95\%$).

\begin{figure}
\includegraphics[width=0.5\textwidth]{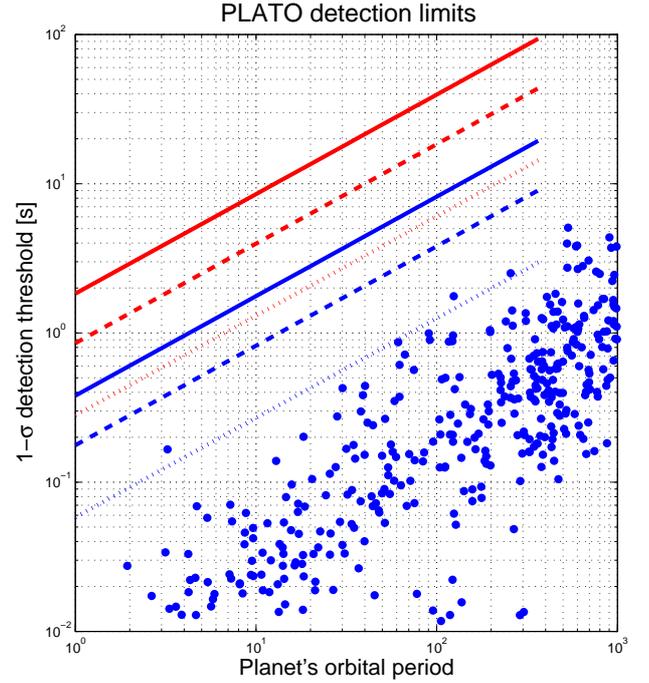}
\caption{Solid lines: expected $1 \sigma$ detection thresholds for PLATO for different scenarios, as a function of the planet's orbital period. Red lines are for Neptune-line planets ($r/R_*=0.035$) and Blue are for Jupiter-like planets ($r/R_*=0.1$), while the solid, dashed and dot-dashed line sets (from top to bottom) are for minimal, good and best-case scenarios, respectively, as described in the text. Dots: $D_{expect}$ of some of the known exoplanets. PLATO planets are likely to have a higher $D_{expect}$ than the current population (see text)}
\label{Limits}
\end{figure}

We compute these detection limits for two planet sizes ($\rho=0.1$ and $\rho=0.035$, Jupiter- and Neptune- like, respectively) orbiting a Sun-like star, and three observational scenarios (see Figure \ref{Limits}): the minimal scenario assumed the P3 star sample requirement of 80 ppm/hr collected over the minimal 2yr (for the long-monitoring phases), the good scenario assumed the P1,P2 star samples requirement of 34 ppm/hr collected over 2.5yr (for the same phases), and the best-case scenario assumed the bright-star ($m_v\leq6$) predicted noise level of 10 ppm/hr collected over (the PLATO goal of) 3yr. One finds that short period giant planets may allow constraining $cos(l)$ to 1sec or better even in the very minimal scenario presented here (minimal noise requirement, minimal observation time span, minimal ingress/egress time). Given the properties of the known systems above, and the predicted PLATO capabilities outlined here, we conclude that the PLATO mission would be able to constrain $cos(l)$ on a large number of planets. Figure \ref{Limits} also includes the calculated $D_{expect}$ value for the known exoplanets -- had they were transiting --  and one can see that the currently-known population lies largely just below PLATO's detection limits. Importantly, since the PLATO target stars are all nearby relative to almost all existing transit surveys, $D_{expect}$ will be, on average, more easily observable in PLATO targets than in the currently known population. Coupled with the large number of targets, we find that a significant number of objects will have an observable $D_{expect}$ using PLATO. We note that much better still detection limits will be availabe for observation of PLATO eclipsing binaries.

We also note that the CHEOPS mission (Broeg et al. 2013) requires a noise level of 150 ppm/min, or $\sim 20$ ppm/hr, for stars brighter than $V\leq9$ and so it may have comparable performance to the PLATO mission (we did not add it to Figure \ref{Limits} to avoid clutter in the figure). The TESS mission is expected to have a performance of $\sim 60$ ppm/hr but different stars will be observed for very different durations (from 27 days to 1 year) so while overall performance will very widely, its best performance will be similar to PLATO's\footnote{No technical publication about TESS is available. We used public data from: \texttt{https://www.youtube.com/watch?v=mpViVEO-ymc}}.

\section{Discussion}

In this letter we have shown that the (cosine of the) on-sky position angle $cos(l)$ can be constrained from high precision transit timing data by comparing the expected and observed parallactic delay if GAIA-class parallax angle to the system is available. Moreover, we have shown that on-sky coplanarity of eclipsing or transiting systems with multiple such transiting components can be probed even if the parallax angle in completely unknown. We note that one a-priory expects that most multi-transiting systems will indeed be well-aligned because misaligned orbits are less likely to exhibit transits of multiple objects.

Knowledge of $cos(l)$ can be useful in a number of ways, for example: (I) in young systems that still have disks one may be able to compare the orientation of the planet and the disk (II) some direct imaging techniques achieve high contrast only on one side of the image plane (e.g. PIAA, Guyon et al. 2012). Knowing $cos(l)$ will allow to aim such an instrument at the most sensitive angle - increasing sensitivity and allowing to not image the other half of the image plane, reducing on-target time. (III) $cos(l)$ may also be measured to change in time, hinting the presence of a multiple systems with strong interaction even if only a single transiting object was previously known. (IV) if the host star (or target binary) has a visual stellar companion, studies on the relation between the orientation of the planetary (binary) orbit and the stellar companion orbit can be made, perhaps pointing to ineraction between the companion and the host natal disk.

Looking forward, we have shown that a large number of known RV-detected planets have parallactic delay in excess of 1 second and so may well have their $cos(l)$ constrained by the future CHEOPS mission. Importantly, the PLATO mission (and obviously the CHEOPS mission) is aimed at a stellar population similar to the RV planet host stars above, giving rise to the expectation that PLATO will measure $cos(l)$ in large scale. By the time they launch, both of these missions will also benefit from high precision parallax measurement for their targets from the GAIA mission, and both of these missions may allow to make the investigations proposed here routine.

\section*{Acknowledgements}

I acknowledges financial support from the Deutsche Forschungsgemeinschaft under DFG GRK. I thank Stefan Dreizler, as well as the anonymous referee, for their useful comments to the manuscript.


\begin{thebibliography}{}

\bibitem[Broeg et al.(2013)]{2013EPJWC..4703005B} Broeg, C., Fortier, A., Ehrenreich, D., et al.\ 2013, European Physical Journal Web of Conferences, 47, 3005 
\bibitem[Borucki et al.(2010)]{2010Sci...327..977B} Borucki, W.~J., Koch,D., Basri, G., et al.\ 2010, Science, 327, 977 
\bibitem[Ford \& Gaudi(2006)]{2006ApJ...652L.137F} Ford, E.~B., \& Gaudi, B.~S.\ 2006, \apjl, 652, L137 
\bibitem[Guyon et al.(2012)]{2012SPIE.8442E..4VG} Guyon, O., Kern, B., Belikov, R., et al.\ 2012, \procspie, 8442,  
\bibitem[Holman \& Murray(2005)]{2005Sci...307.1288H} Holman, M.~J., \& Murray, N.~W.\ 2005, Science, 307, 1288 
\bibitem[Ohta et al.(2005)]{2005ApJ...622.1118O} Ohta, Y., Taruya, A., \& Suto, Y.\ 2005, \apj, 622, 1118
\bibitem[Tregloan-Reed et al.(2013)]{2013MNRAS.428.3671T} Tregloan-Reed, J., Southworth, J., \& Tappert, C.\ 2013, \mnras, 428, 3671 
\bibitem[Rauer \& Catala(2011)]{2011IAUS..276..354R} Rauer, H., \& Catala, C.\ 2011, IAU Symposium, 276, 354 


\end{thebibliography}
\end{document}